\documentclass[12pt]{article}
\usepackage{epsfig,latexsym}

\input epsf
\begin{document}
\font\cmss=cmss10 \font\cmsss=cmss10 at 7pt
\hfill CPHT-S060.12.01
\par\hfill Bicocca-FT-01-31
\par\hfill IFUP-TH 2001/42
\par\hfill IC/2001/173\\
\vskip .1in \hfill hep-th/0112236

\hfill

\vspace{20pt}

\begin{center}
{\Large \textbf{Some Comments on}}
{\Large \textbf{$\mathcal{N}=1$ Gauge Theories from Wrapped Branes}}
\end{center}

\vspace{6pt}

\begin{center}
\textsl{R. Apreda ${}^{a}$, F. Bigazzi ${}^{b}$, A. L. Cotrone ${}^{c}$, M. Petrini ${}^{d}$ and A. Zaffaroni ${}^{c}$} \vspace{20pt}

\textit{a) Dipartimento di Fisica, Universit\`{a} di Pisa, via Buonarroti, 2; I-56127 Pisa, Italy.}\\
\textit{b) The Abdus Salam International Centre for Theoretical Physics, Strada
Costiera, 11; I-34014 Trieste, Italy.}\\
\textit{c) Dipartimento di Fisica, Universit\`{a} di Milano-Bicocca, P.zza della Scienza, 3; I-20126 Milano, Italy.}\\
\textit{d) Centre de Physique Th\'eorique, \`Ecole Polytechnique, 48 Route de Saclay; F-91128 Palaiseau Cedex, France.}\\

\end{center}

\vspace{12pt}

\begin{center}
\textbf{Abstract }
\end{center}

\vspace{4pt} {\small \noindent
We discuss various aspects of gauge theories realized on the
world-volume of wrapped branes. In particular we analyze the coupling of
SYM operators to space-time fields both in $\mathcal{N}=1$ and $\mathcal{N}=2$ models
and give a description of the gluino condensate
in the Maldacena-Nu\~nez $\mathcal{N}=1$ solution. We also explore the seven-dimensional
BPS equations relevant for these solutions and their generalizations.
}
\vfill
\vskip 5.mm
 \hrule width 5.cm
\vskip 2.mm
{\small
\noindent
apreda@df.unipi.it\\
bigazzif@ictp.trieste.it\\
aldo.cotrone@mib.infn.it\\
Michela.Petrini@cpht.polytechnique.fr\\
alberto.zaffaroni@mib.infn.it}

\newpage
\section{Introduction}
$\mathcal{N}=1$ and $\mathcal{N}=2$ SYM theories can be obtained as the low energy limit
of wrapped five-branes in Type IIB superstring theory. In both cases, the
five-branes are wrapped on a supersymmetric cycle of the ambient
geometry and  a partial twist \cite{vafa,mn1} of the world-volume
theory is necessary to preserve some supersymmetry.
The amount of
surviving supersymmetry depends on how the cycle is embedded in the
ambient geometry. Information about the quantum field theory can be 
obtained with a geometric engeenering approach \cite{lerche}.
In AdS/CFT context, solutions for both $\mathcal{N}=1$ and $\mathcal{N}=2$ pure SYM in four dimensions with gauge group $SU(N)$ have been
found by exploiting seven dimensional gauged supergravity
\cite{mn2,martelli,us}. The $\mathcal{N}=2$ solution correctly reproduces
the one-loop $\mathcal{N}=2$ effective action but it is singular at the scale
where, in QFT, instantonic corrections take over. This IR singularity
could be possibly solved by some modifications of the enhan\c con mechanism \cite{jpp}. On the other hand,
the $\mathcal{N}=1$ Maldacena-Nu\~nez (MN) solution, upon inclusion of non-abelian background fields, is completely regular.

In this note, we address some aspects of these constructions.
All solutions are asymptotic to the linear dilaton background, which
is holographically dual to the little string theory associated with the
five-branes. It makes sense then
to ask whether and to which extent the AdS/CFT rules apply to such
solutions. We will show that one can make a reasonable identification
of dual operator-fields and asymptotic behaviors. In particular,
in the $\mathcal{N}=1$ case, we find that the space-time deformation which
de-singularizes the MN solution is dual to the gaugino condensate.
The asymptotic behavior of the latter can be explicitly computed and
it exhibits the expected dependence on the QFT parameters. Our results
closely parallels those found for the Klebanov-Strassler (KS)
solution \cite{ks,sonne}. We also classify all operators that are
involved in generalized solutions such as, for example,
the $\mathcal{N}=2\rightarrow \mathcal{N}=1$ breaking. Some relevant BPS equations for this case
are collected in the Appendices, leaving the explicit solution
for future work. Finally, we also explicitly write and solve
 the BPS equations for the MN solution in seven dimensions.
They are written and solved in many papers in literature, but,
curiously, never in the
natural setting, which is seven dimensional gauged supergravity.
\section{The relevant operators}
We consider the general setting of wrapped five-branes with
world-volume $R^4\times S^2$, which can be
easily adapted to describe both $\mathcal{N}=1$ and $\mathcal{N}=2$ theories.
In the case of N flat NS5 branes the world-volume theory is a (1,1)
six-dimensional SYM. The
$SO(4)_R$ R-symmetry acts on the transverse space $R^4$, whose directions
we label with an index $i=1,...,4$. As well known, there are no
covariantly constant spinors on $S^2$.
Some supersymmetry is preserved only
if an abelian background field in $SO(4)$ is turned on in order to
cancel the spin connection on $S^2$ \cite{mn1}.
From the schematic formula for the variation of a fermion:
\begin{equation}
\delta \Psi\sim D_{\mu}\epsilon=(\partial_{\mu}+\omega_{\mu}^{\nu\rho}\gamma^{\nu\rho}
-A_{\mu}^{ij}\Gamma^{ij})\epsilon,
\label{twist0}
\end{equation}
we see that the surviving spinors are those satisfying the twist condition:
\begin{equation}
(\omega_{\mu}^{\nu\rho}\gamma^{\nu\rho}
-A_{\mu}^{ij}\Gamma^{ij})\epsilon=0.
\label{twist}
\end{equation}
The relevant $U(1)$ fields
are those of the decomposition
$SO(4)_R\rightarrow U(1)_{(1)}\times U(1)_{(2)}$, $A_{(1)}$
and $A_{(2)}$ .
In the $\mathcal{N}=2$ solution \cite{martelli,us}
only $A_{(1)}$ is turned on,
while in the $\mathcal{N}=1$
MN solution \cite{mn2}, both $A_{(1)}$ and $A_{(2)}$ are turned on.

Let us discuss first the $\mathcal{N}=2$ case.
The relevant type IIB space-time fields are described by the $SO(4)_R$
gauged supergravity in seven dimensions \cite{sz}. The bosonic Lagrangian
and the fermionic shifts are written in Appendix A.
The gauged supergravity has a vacuum corresponding to a set of
coincident five-branes.
The theory contains ten scalar fields parameterizing a symmetric
matrix $T_{ij}$.
Before twisting, they are dual to the bilinear operator
${\mbox Tr}X_iX_j$, where $X_i$
are the six dimensional scalar fields. Upon compactification
on $S^2$ and related twist, $X_1$ and $X_2$ get masses.
The massless complex scalar $\phi=X_3+iX_4$ parameterizes the $\mathcal{N}=2$
moduli space.
The $\mathcal{N}=2$ solutions that have been found in \cite{martelli,us} always involve a diagonal $T_{ij}$ and only three scalars. In seven dimensions
they have the following form \cite{martelli,us}:
\begin{eqnarray}
ds_{7}^2 &=&e^{2f}(dx_{4}^2 + d\rho^2) + e^{2g}(d\theta^2 + \sin^{2}{\theta} d\phi^2),\nonumber \\
A_{(1)}&=&{1\over 2}\cos\theta d\phi,\nonumber \\
T_{ij}&=&{\mbox {diag}}(e^{2\lambda_1}, e^{2\lambda_1},
e^{2\lambda_2}, e^{2\lambda_3}),
\label{o1}
\end{eqnarray}
where $g$,$f$ and $\lambda_i$ are functions of the radial coordinate $\rho$ \cite{martelli,us}. Moreover these 7$d$ solutions
can be explicitly lifted to ten dimensions \cite{martelli,us}.
The effective action for a probe in these backgrounds can be written in
manifest $\mathcal{N}=2$ language with coupling constant:
\begin{eqnarray}
{\mbox {Solution}} &I&\,\,\,\tau(z)={Ni\over \pi}\log {z\over \Lambda},
\nonumber\\
{\mbox {Solution}} &II&\,\,\,\tau(z)={Ni\over \pi}\left ({\mbox {Arcosh}}{z\over 2b}+ {\mbox {const}}\right ).
\label{tau}
\end{eqnarray}
Solution $I$ corresponds to $\lambda_2 = \lambda_3$, while in
Solution $II$ all the three scalars are different.
As discussed in \cite{us,evans}, the presence of $\lambda_2,\lambda_3$
indicates that the operators ${\mbox Tr} \phi\bar\phi$
and ${\mbox Tr} \phi^2$ are turned on. These solutions indeed describe
points in the Coulomb branch of the $\mathcal{N}=2$ theory.
${\mbox Tr} \phi\bar\phi$
and ${\mbox Tr} \phi^2$ are dual to $\lambda_2+\lambda_3$ and  $\lambda_2-\lambda_3$ respectively. This is consistent with the fact that the rotationally invariant Solution $I$ has $\lambda_2=\lambda_3$ \cite{us}.
A detailed discussion
of the asymptotic behavior of the fields can be found in \cite{evans}.

Let us consider the fermionic fields.
The six dimensional theory
has fermions $\Psi=\Psi^+ +\Psi^-$ transforming in the
representation $(4,2)+(4^\prime,2^\prime)$
of $SO(5,1)\times SO(4)_R$.  We have $\Psi^{\pm}=\pm\gamma_7\Psi^{\pm}$.
We choose the following basis
for the six dimensional gamma matrices
\begin{equation}
\gamma^{\mu}=\gamma_{(4)}^{\mu}\otimes 1, \qquad \gamma^{\theta,\phi,7}=\gamma_{(4)}^5 \otimes
\sigma^{1,2,3},
\label{o3}
\end{equation}
with conjugation matrix $C^{(6)}=C^{(4)}\otimes \sigma_2$.
We write the $SO(4)= SU(2)^+ \otimes SU(2)^-$ action on spinors
in a basis of sigma matrices for $\Psi^{\pm}$,
\begin{equation}
\Gamma^{12}\pm \Gamma^{34}=2i\sigma_3^{\pm},\,\,\,\,\Gamma^{24}\pm \Gamma^{31}=2i\sigma_1^{\pm},\,\,\,\,\Gamma^{14}\pm
\Gamma^{23}=-2i\sigma_2^{\pm}.
\end{equation}
Notice that the previously defined $A_{(1)}$ and $A_{(2)}$ are related to
the $U(1)$ subgroups of $SU(2)^\pm$ by a change of basis.
Since both $SO(5,1)$ and $SO(4)_R$ act on spinors, a symplectic-Majorana
condition can be imposed on the fermions, reducing the supersymmetry to
the expected 16 real supercharges.
 The
symplectic-Majorana condition reads $\Psi^\alpha=C_{(6)}\gamma_0^T
\Omega^{\alpha\beta}\Psi^{*\beta}$, where the
symplectic matrix is $\Omega^{\pm}=i\sigma_2^{\pm}$ for each
$SU(2)^{\pm}$ group. It is convenient to write the
six dimensional spinors as two-by-two matrices
on which the Lorentz $\sigma$'s
(associated with the spin connection on $S^2$)
act on the left and the gauge $\sigma^{\pm}$ on the right.
With these conventions, the six dimensional fermions are given by:
\begin{eqnarray}
\Psi=\Psi^+ \oplus \Psi^-, \qquad \Psi^+=\left( \begin{array}{cc}
p & q \\ iq^c & -ip^c \end{array} \right), \qquad \Psi^-=\left(
\begin{array}{cc} \tilde{p} & \tilde{q} \\ i\tilde{q}^c &
-i\tilde{p}^c \end{array} \right).\label{psi}
\end{eqnarray}
From $\gamma^7=\gamma_{(4)}^5\otimes\sigma_3$ we see that
$p$ and $q$ are four dimensional Weyl spinors of positive chirality
while $\tilde p$ and $\tilde q$ are four dimensional Weyl spinors of
negative chirality.
\vskip 0.5truecm
\begin{center}
{\small
\begin{tabular}{ccc|cccccccccccc}
 & & & & & & & & & &   \\[-1mm]
 & & & & $p=\lambda$ &  & $\tilde{p}=\bar{\psi}$ &  & $q$ &  & $\tilde{q}$  \\[-1mm]
 & & & & & & & & & &  \\[-1mm]
\hline
 & & & & & & & & & &   \\[-1mm]
 & $U(1)_R=U(1)_{(2)}$ & & & 1 &  & -1 &  & -1 &  & 1  \\[-1mm]
 & & & & & & & & & & & & & &  \\[-1mm]
 & $U(1)_J=U(1)_{S^2}$ & & & 1 &  & 1 &  & 1 &  & 1  \\[-1mm]
 & & & & & & & & & &  \\[-1mm]
 & $U(1)_{(1)}$ & & & 1 &  & 1 &  & -1 &  & -1  \\[-1mm]
 & & & & & & & & & &   \\
\end{tabular}}
\vskip 0.3truecm
Table 1: Charge assignment of the spinors.
\end{center}
\vskip 0.5truecm
The twisted $U(1)$ in the $\mathcal{N}=2$ solution is $\Gamma^{12}=i(\sigma_3^+ +
\sigma_3^-)$ and the twist condition reads:
\begin{equation}
i\sigma_3^+\Psi^+=\gamma^{\theta\phi}\Psi^+,\,\,\,\, i\sigma_3^-\Psi^-=\gamma^{\theta\phi}\Psi^-.
\label{twistn2}
\end{equation}
The spinors $p$ and $\tilde p$, which have the same charges under
the twisted $U(1)$ and the $S^2$ generator $\gamma^{\theta\phi}$,
remain massless in four dimensions. Analogously $q$ and $\tilde q$ are
massive four dimensional fields.
The four dimensional theory has an $SU(2)_R\times U(1)_R$
R-symmetry.
The two chiral spinors in the $\mathcal{N}=2$ vector multiplet,
$\psi$ and $\lambda$ (this one being the gaugino), have different
charges under $U(1)_R$ and $U(1)_J$ (the abelian generator of
$SU(2)_R$) \cite{seibergwitten}. While $\psi$ has charges $+1$ and
$-1$ respectively, $\lambda$ has charge $+1$ under both of them.
In the supergravity language $U(1)_R$ is the untwisted $U(1)_{(2)}$
\cite{martelli}\cite{us},
so in our notation it is generated by $\Gamma^{34}=i(\sigma_3^+ -
\sigma_3^-)$. On the other
hand $U(1)_J$ is the isometry of the sphere on which we are
wrapping the branes, so it is generated by $\gamma^{\theta\phi}$.
Now, it is easy to verify that $p$ and $\tilde{p}$ have both
charge $+1$ with respect to $\gamma^{\theta\phi}$, and
charges $+1$ and $-1$ respectively under $\Gamma^{34}$,
so that we can safely identify them with $\lambda$ and $\bar\psi$.

We are interested in two special operators, the bilinears
$\lambda\lambda$ and $\psi\psi$.
They can be easily obtained
from the coupling:
\begin{equation}
A^\mu_{ij}\bar{\Psi}\gamma^\mu\Gamma^{ij}\Psi.\label{apsipsi}
\end{equation}
We choose the combination of gauge fields:
\begin{eqnarray}
A_\theta & = & {1\over 2} a\,\eta_{2}^{+}+{1\over 2}\tilde{a}\,
\eta _{2}^{-},\nonumber \\
A_\phi & = & {1\over 2}a \sin{\theta}\, \eta _{3}^{+}+{1\over
2}\tilde{a} \sin{\theta}\, \eta _{3}^{-}. \label{A}
 \label{A1}
\end{eqnarray}
The $\eta$ matrices are the generators of the $SU(2)^{\pm}$ in
the $SO(4)$ notation and take the form:
\begin{eqnarray}
\eta _{1}^{\pm}={1\over 2} \texttt{\footnotesize $\left( \begin{array}{cccc} 0 & 1 & 0 & 0
\\ -1 & 0 & 0 & 0 \\ 0 & 0 & 0 & \pm 1 \\ 0 & 0 & \mp 1 & 0 \end{array}
\right)$}\,\, \eta_{2}^{\pm}={1\over 2}  \texttt{\footnotesize $
\left( \begin{array}{cccc} 0 & 0 & \mp 1 & 0 \\ 0 & 0 & 0 & 1 \\
\pm 1 & 0 & 0 & 0 \\ 0 & -1 & 0 & 0
\end{array} \right)$}\,\,
\eta _{3}^{\pm}={1\over 2}  \texttt{\footnotesize $\left(\begin{array}{cccc} 0 & 0 & 0 & 1 \\
0 & 0 & \pm 1 & 0 \\ 0 & \mp 1 & 0 & 0 \\ -1 & 0 & 0 & 0 \end{array}
\right)$}
\label{eta}
\end{eqnarray}
The calculation is straightforward. The $q$
and $\tilde{q}$ spinors get opposite contributions from the
$\theta$ and $\phi$ components of (\ref{apsipsi}) and they do not
survive in the reduction. We then find that $a$ and $\tilde a$ couple
to fermionic bilinears:
\begin{equation}
a\,\bar{\lambda^c}\lambda, \qquad \tilde{a}\bar{\psi^c}\psi.
\label{bili}
\end{equation}
These two fermionic bilinears are necessary ingredients for the
soft breaking $\mathcal{N}=2\rightarrow \mathcal{N}=1$ and for the study of the
gaugino condensate in $\mathcal{N}=1$ theories.
\section{The gaugino condensate}
The previous formalism is easily adapted to $\mathcal{N}=1$ theories and the MN
solution. The relevant twisted $U(1)$ field is now $\sigma_3^+=
-i(\Gamma^{12}+\Gamma^{34})/2$, so that both $A_{(1)}$ and $A_{(2)}$
are turned on. The twist condition is now:
\begin{equation}
i\sigma_3^+\Psi^+=\gamma^{\theta\phi}\Psi^+.
\label{twistn22}
\end{equation}
From this equation (see also Table 1), we easily see that
the only massless fermionic field  is $p\sim\lambda$.
The MN solution is reviewed in Appendix B. The fields that are turned on
are the metric (then the fields $f,g$ of equation (\ref{o1})),
the non-abelian gauge fields:
\begin{equation}
A = {1\over 2} [\cos{\theta}\,d\phi \, \eta _{1}^{+} + a(\rho)\, d\theta \, \eta
_{2}^{+}+ a(\rho)\, \sin{\theta}\,  d\phi \,  \eta _{3}^{+}],
\label{mn1}
\end{equation}
and the singlet scalar $\lambda=\lambda_1=\lambda_2=\lambda_3$.
The solution reads:
\begin{equation}
e^{2h}=\rho \coth{2\rho}-{\rho^2 \over \sinh^2{2\rho}}-{1\over4},\qquad
a={2\rho \over \sinh{2\rho}},\qquad
e^{10\lambda}={2e^h\over \sinh{2\rho}},
\end{equation}
where $h=g-f$. The ten dimensional
solution describing N NS five-branes in the string frame is:
\begin{equation}
ds_{st}^2=dx_{(4)}^2+\alpha^\prime N [d\rho^2+e^{2h(\rho)}(d\theta^2 + \sin^{2}{\theta} d\phi^2)+ ds^2_{S^3}].
\label{stringmetric}
\end{equation}
This solution is an appropriate description in the UV ($\rho\gg 1$).
In the IR the appropriate description is the one with D5-branes \cite{mn2}. In the UV the non-abelian gauge fields are asymptotically zero, and the behavior of the MN solution
can be described by the following simpler
solution of the BPS equations with
only the abelian component of the gauge fields (the one defining
the twist):
\begin{equation}
e^{2h}=\rho, \qquad 5\lambda={1\over 4}\log{\rho} - \rho.
\end{equation}

We are now ready to discuss the gaugino condensate in the MN solution.
We see from formula (\ref{bili}) that the particular combination of
non-abelian gauge fields $a$ in the MN
solution is dual to the gaugino condensate. It is remarkable that
the space-time field
that de-singularizes the solution is associated with the condensate,
i.e. with the non-trivial IR dynamics of the $\mathcal{N}=1$
SYM theory. A closely related situation has been noticed for
the KS solution \cite{bgz,sonne}.

By adapting AdS/CFT rules,
we can determine the value of the gaugino condensate from the
asymptotic behavior of the dual supergravity field.
We face various conceptual problems. The decoupling of SYM scale
from the six-dimensional little string scale requires to go beyond
the supergravity approximation \cite{mn2}. Moreover, for related reasons,
the identification of the radial
parameter with the energy scale in the MN solution is difficult and ambiguous
\cite{pp,mn2}.
We nevertheless obtain a reasonable result by using a Born-Infeld analysis.
From the BI action for a stack of D5-branes wrapped over
a two-sphere we determine the four dimensional gauge coupling:
\begin{equation}
{1\over g_{YM}^2}= {\tau_5\over 2}(2\pi\alpha')^2\int_{S^2} e^{-\Phi}\sqrt{g_{\theta\theta}g_{\varphi\varphi}},
\end{equation}
where $\tau_5= (2\pi)^{-5}(\alpha')^{-3}$ and where we
used the conventions (see for example \cite{peskin}):
\begin{equation}
L= - {1\over 4g_{YM}^2}(F^a_{\mu\nu})^2= -{1\over 2g_{YM}^2}Tr(F_{\mu\nu}F^{\mu\nu}).
\end{equation}
From the UV behavior of the metric and the dilaton for the D5 solution
(perform an S-duality in (\ref{stringmetric}))
we can thus read the UV behavior of the gauge coupling:
${1\over g_{YM}^2}= {N\over4\pi^2}\rho$,
which, compared with the gauge theory result
${1\over g_{YM}^2}= {3N\over 8\pi^2}\log(\mu/\Lambda)$,
gives the radius/energy relation:
\begin{equation}
\rho={3\over2}\log(\mu/\Lambda).
\label{raden}
\end{equation}
The field $a$ vanishes in the UV as $a\sim \rho e^{-2\rho}$, that is, from (\ref{raden}), as:
\begin{equation}
a\sim \left ({\Lambda^3\over \mu^3}\right )\log {\mu\over \Lambda}.
\label{mn4}
\end{equation}
This result has the expected dependence on the scale $\Lambda$, since
the gaugino condensate is an operator with (protected) dimension three.
Moreover, under the chiral symmetry $Z_{2N}$, which is a subgroup of
$U(1)_R$ \cite{mn2}, the condensate transforms with a phase,
as expected~\footnote{
The gaugino condensate is BPS related to the tension of a domain wall,
which can be easily computed in this model \cite{sonne} using the IR form
of the metric. It is however difficult to compare the two results.}.
A similar result was found for the KS solution \cite{sonne}
\footnote{In that case, the logarithmic correction was associated with the
logarithmically varying AdS radius in the UV. In our case, the
asymptotic solution should decompactify to six-dimensions. The fact that
we nevertheless get a sensible four dimensional
result could be related to a topological nature of the twisted theory
on the two-sphere \cite{vafa}.}.
In the AdS/CFT philosophy \cite{kw}, the two independent
solutions of the asymptotic second order equations of motion for a field
are associated in QFT with a deformation by the dual operator
(the dominant solution) and with a different vacuum of
the theory where the dual operator has a VEV
(the sub-dominant
solution). Using the explicit second order equations for $a$ in the
asymptotic background, written for example in \cite{gubser}, one can
explicitly check that the BPS equations single out the behavior of $a$
appropriate for a condensate.
\section{Conclusions}
The analysis in Section 2 is useful for identifying the fields that
are needed for more general solutions, such as the
breaking $\mathcal{N}=2\rightarrow \mathcal{N}=1$. We expect that both $SU(2)^{\pm}$
non-abelian gauge fields are turned on, as in formula (\ref{A}),
$a$ representing the gaugino condensate and $\tilde a$ the soft
breaking mass term for fermions. The corresponding mass term for the
scalars should be $(\lambda_2+\lambda_3)/2$.
We can expect  both operators ${\mbox Tr}\phi^2$ and
${\mbox Tr}\phi\bar\phi$ to have a VEV \cite{seibergwitten,ds} in QFT.
On the supergravity side this should correspond to a solution
where all the fields $(\lambda_1,\lambda_2,\lambda_3,a,\tilde a)$
are turned on.
The BPS equations for this case are written in
Appendix A. We leave their resolution for future work.
It is not clear that a solution exists, because there are more
equations than independent variables. Since we cannot rely on any
consistent truncation argument, a more general ansatz for the scalar
and gauge fields could be required.
We notice, however, that a simplified model with $\lambda_2=\lambda_3$
and $\beta=0$ admits a solution, despite the fact that the number of
equations is redundant.
Even if this solution is IR singular, we could interpret its
existence as a signal for the existence of the (expected) more
general one. Details are presented in Appendix C.
\vskip .1in
\noindent \textbf{Acknowledgments}
\vskip .1in \noindent
We would like to thank N. Evans for useful discussions.
R. A., F. B., A. L. C. and A. Z. are partially supported by INFN. R. A., A. L. C. and A. Z. are partially supported by MURST and
the European Commission TMR program HPRN-CT-2000-00131,
wherein they are associated to the University of Padova. M.P. is
partially supported by the European Commission TMR program
HPRN-CT-2000-00148. R. A. would like to thank the physics department of Milano-Bicocca University for kind hospitality.
\section*{Appendix A: General equations}
The relevant supergravity is the seven dimensional $SO(4)$ gauged
supergravity \cite{sz}. It can be obtained by the maximally
supersymmetric $SO(5)$ gauged supergravity performing a suitable
singular limit on the scalars \cite{cve1,us}.
We use the notation of
\cite{minasian}. The relevant terms of the bosonic lagrangian are:
\begin{equation}
2\kappa^2e^{-1}{\cal L} = R + {1\over 2} m^2 (T^2 - 2T_{ij}T^{ij}) -
Tr(P_{\mu} P^{\mu}) - {1\over 2} (V_I{}^i V_J{}^j F_{\mu \nu}^{IJ})^2.
\end{equation}
In the expression above, the signature of space-time is mostly
plus, and $I$ and $i$ are respectively the gauge and composite $SO(4)$
indices.
$V_i^I$ is the symmetric matrix for the ten scalar degrees of freedom
parameterizing the $SL(4,R)/SO(4)$ coset space, and the $T$ matrix is
defined as $T_{ij} = V_i^{-1\,I}
V_j^{-1\,J} \delta_{IJ}$,  $T=T_{ij}\delta_{ij}$. The kinetic term for the scalars $P_{\mu}$ is the symmetric part of $V_i^{-1\,I} {\cal D}_{\mu} V_I{}^j = \left( Q_{\mu} \right)_{[ij]} + \left( P_{\mu}
\right)_{(ij)}$, where the
covariant derivatives are defined as ${\cal D}_\mu V_I{}^j=\partial_\mu
V_I{}^j+2m A_{\mu\, I}^J V_J{}^j$ on the scalars, ${\cal D}_\mu \psi=(\partial_{\mu}+{1 \over 4}Q_{\mu ij}\Gamma^{ij}+{1 \over 4}\omega_{\mu}^{\nu\lambda}\gamma^{\nu\lambda})\psi$ on the spinors.

With an $SO(4)_R$ gauge rotation,
$T_{ij}$ can be always diagonalized.
In this appendix we write down the general formulae for the
supersymmetry variations of fermions with only three diagonal scalars:
\begin{equation}
V_I{}^i = {\rm diag}( e^{-\lambda_1},  e^{-\lambda_1},
e^{-\lambda_2},  e^{-\lambda_3}).
\end{equation}
We take the $SO(4)=SU(2)^+ \times SU(2)^-$ gauge fields of the form:
\begin{eqnarray}
A & = & \alpha \,[\cos{\theta} \, d\phi \, \eta _{1}^{+} + a(\rho) \,d\theta \, \eta
_{2}^{+}+ b(\rho) \, \sin{\theta} \, d\phi \, \eta _{3}^{+}]+ \nonumber \\
&& \beta \, [\cos{\theta}
\, d\phi \, \eta _{1}^{-} + \tilde{a}(\rho) \, d\theta \, \eta _{2}^{-} +
\tilde{b}(\rho) \, \sin{\theta} \, d\phi \, \eta _{3}^{-}] \label{AA}.
\end{eqnarray}
The field strength is normalized as $F=dA+2m[A,A]\,$ and
the $\eta$ matrices are given in (\ref{eta}).
The ansatz for the metric (in the Einstein frame) is:
\begin{equation}
ds_{7}^2 = e^{2f}(dx_{4}^2 + d\rho^2) + e^{2g}(d\theta^2 + \sin^{2}{\theta} d\phi^2).
\label{m2}
\end{equation}
In this paper we do not use different notations for curved and flat
indices. To  pass from the formers  to the latters we must multiply
$\gamma^\phi,\gamma^\theta,\gamma^{\rho,\chi}$ by the inverse vielbein
($\chi=0,1,2,3$ labels the four dimensional coordinates).
From (\ref{m2}) it follows that the non trivial components of the spin
connection are:
\begin{equation}
\omega_{\chi}^{\chi\rho}= f', \quad \omega_{\theta}^{\theta\rho}=g'e^{g-f}, \quad \omega_{\phi}^{\phi\rho}=g'e^{g-f}\sin{\theta},  \quad \omega_{\phi}^{\phi\theta}=\cos{\theta},
\label{spinconn}
\end{equation}
where the prime denotes a derivation with respect to $\rho$.\\
The general form of the supersymmetry variations can be found in \cite{minasian}:
\begin{eqnarray}
\delta\psi_\mu &=& \left[ {\cal D}_\mu+
{1\over4} \gamma_\mu\gamma^\nu V_i^{-1\,I} {\partial}_{\nu} V_I{}^i
+{1\over 4} \Gamma^{ij}F_{\mu\lambda}^{ij}\gamma^\lambda
\right] \epsilon,\nonumber\\
\delta(\Gamma^{\hat{i}}\lambda_{\hat{i}}) &=& \left[ {m \over 2}(T_{\hat{i}j}-{1\over 5} T \delta_{\hat{i}j})\Gamma^{\hat{i}j}+{1\over2}\gamma_\mu P^\mu_{\hat{i}j} \Gamma^{\hat{i}j}+{1\over 16}\gamma^{\mu\nu}(\Gamma^{\hat{i}}\Gamma^{kl}\Gamma^{\hat{i}}-{1\over 5}\Gamma^{kl})F_{\mu\nu}^{kl}
\right] \epsilon.\qquad
\end{eqnarray}
with $F_{\mu\lambda}^{ij}=V_I{}^iV_J{}^jF_{\mu\lambda}^{IJ}$. Notice that the index $\hat{i}$ is not summed over.
Being the spinor $\epsilon$ charged under $SU(2)^+\times SU(2)^-$,
we separate the two components in $\epsilon=\epsilon^+ \oplus
\epsilon^-$. If we want to preserve $\mathcal{N}=1$ supersymmetry, we can concentrate only on $\epsilon^+$.
There are some constraints which come from the dependence on
$\theta$ in the fermionic variations.
From the ${\cos\theta \over \sin\theta}$ terms in the gravitino
shift, one gets $b=2\,m\,\alpha\,a$ and $\tilde{b}=2\,m\,\beta\,\tilde{a}$.
The contribution in $\cos\theta$ in the gravitino shift gives the twist condition:
\begin{equation}
\left[
\gamma^{\phi\theta}+m[(\alpha+\beta)+{1\over 2}(\alpha-\beta)(e^{\lambda_2-\lambda_3}+e^{\lambda_3-\lambda_2})]i\sigma_3^+\right]\epsilon^+=0.\label{Atwist0}
\end{equation}
Eq.(\ref{Atwist0}) takes a more illuminating form in the two interesting
cases, a) pure $\mathcal{N}=1$ with $\beta=0$ and
$\lambda_2=\lambda_3=\lambda_1$ (MN)
and b) $\mathcal{N}=2\rightarrow \mathcal{N}=1$ breaking with
$\alpha=\beta$:
\begin{equation}
\left[
\gamma^{\phi\theta}+2m\alpha i\sigma_3^+\right]\epsilon^+=0,\label{Atwist}
\end{equation}
which implies $2m\alpha=\pm 1$, for consistency.

Finally, the remaining parts of the fermionic variations only depend on
functions of $\rho$ and
we obtain the following BPS equations:
\begin{eqnarray}
\delta\psi_{\chi} &\rightarrow& f'+x'
=0,\nonumber\\
\delta\psi_\rho &\rightarrow& \left[\partial_{\rho}+{1\over2}x'+{1\over2}e^{-h}\gamma^\theta
i\sigma_1^+(a'\cosh{z}+\tilde{a}'\sinh{z})\right]\epsilon^+=0,\nonumber \\
\delta\psi_\phi &\rightarrow& \left[ h'e^h+\gamma^{\rho\theta}
i\sigma_1^+(a\cosh{z}\cosh{y}+\tilde{a}\sinh{z}\sinh{y})
+{1\over2}\gamma^\theta
i\sigma_1^+(a'\cosh{z}+\tilde{a}'\sinh{z})\right. + \nonumber \\
&& \left. -{1\over2}e^{-h}\gamma^\rho[(a^2-1)\cosh{y}-(\tilde{a}^2-1)\sinh{y}]\right]\epsilon^+=0,\nonumber \\
\delta\lambda_i &\rightarrow& \left[e^{-y}\sinh{2z}-z'\gamma^\rho+\gamma^\theta
i\sigma_1^+e^{-h}(a\sinh{z}\cosh{y}+\tilde{a}\cosh{z}\sinh{y})\right.+ \nonumber \\
&&\left.+{1\over2}\gamma^{\rho\theta}
i\sigma_1^+e^{-h}(a'\sinh{z}+\tilde{a}'\cosh{z})\right]\epsilon^+=0,\nonumber \\
&& \left[ {1\over5}(e^y+e^{-y}\cosh{2z})+{1\over10}e^{-2h}[(a^2-1)\cosh{y}-(\tilde{a}^2-1)\sinh{y}]\right. + \nonumber \\
&&\left. -x'\gamma^\rho-{1\over5}\gamma^{\rho\theta}
i\sigma_1^+e^{-h}(a'\cosh{z}+\tilde{a}'\sinh{z})\right]\epsilon^+=0,\nonumber \\
&& \left[(e^y-e^{-y}\cosh{2z})-{1\over2}e^{-2h}[(a^2-1)\sinh{y}-(\tilde{a}^2-1)\cosh{y}]\right. + \nonumber \\
&&\left. -y'\gamma^\rho+2\gamma^{\theta}
i\sigma_1^+e^{-h}(a\cosh{z}\sinh{y}+\tilde{a}\sinh{z}\cosh{y})\right]\epsilon^+=0,
\end{eqnarray}
where $x=\lambda_1 +{\lambda_2 +\lambda_3 \over 2}$, $y=\lambda_1 -{\lambda_2 +\lambda_3 \over 2}$ , $z={\lambda_2 -\lambda_3 \over 2}$ and $h=g-f$ (the equation for the $\theta$ component of the gravitino shift
 is equivalent to that for $\phi$).
Notice that there are more equations than independent fields.
Since we cannot rely on any
consistent truncation argument, a more general ansatz for the scalar
and gauge fields could be required.
\section*{Appendix B: the Maldacena-Nu\~nez solution}
Let us review first the singular solution of \cite{mn2}, that has
only one scalar and gauge group $U(1)^+$. We have to put
$\lambda_1=\lambda_2=\lambda_3=\lambda$, $a=b=\beta=0$ in the equations.
We can take the explicit representation for the $\gamma$ matrices:
$\gamma^{\theta,\phi,\rho}=\gamma_5 \otimes \sigma^{1,2,3}$; the
symplectic-Majorana spinor $\epsilon^+$ will be as in (\ref{psi}),
but now $p$ and $q$ are Dirac spinors and functions of $\rho$. The twist (\ref{Atwist}) in this representation gives $\alpha=1/2$, $q=0$ ($m=1$ in our units).
The only non-vanishing equations are:
\begin{equation}
h'={1\over2}e^{-2h},\qquad \qquad
\lambda '=-{1\over 5}+{1\over 20}e^{-2h}.\label{u11}
\end{equation}
The solution is given in \cite{mn2} ($\lambda$ is related to the dilaton
of reference \cite{mn1} by $\phi=5\lambda$):
\begin{equation}
e^{2h}=\rho, \qquad \qquad 5\lambda={1\over 4}\log{\rho} - \rho.
\end{equation}
The non-singular solution, corresponding to one scalar and group $SU(2)^+$, can be obtained from the general equations letting $\lambda_1=\lambda_2=\lambda_3=\lambda$ and $\beta=0$. In this case the structure of the equations is no more linear. In fact, with the definitions:
\begin{equation}
A={1\over2}h'e^h, \quad B={1\over2}a, \quad C={1\over4}e^{-h}(a^2-1), \quad D=-{1\over4}a',
\end{equation}
the $\psi_\theta$ equation becomes:
\begin{equation}
[\gamma^{\theta\rho}A+iB\sigma_1^+ +i\gamma^\phi C\sigma_3^+ +i\gamma^\rho D\sigma_1^+]\epsilon^+=0,
\label{pippo}\end{equation}
that can be rewritten as:
\begin{equation}
i\sigma_1^+\gamma^{\rho\theta}\epsilon^+=(\Delta + \Pi \gamma^\rho)\epsilon^+,\label{ventitre}
\end{equation}
with:
\begin{equation}
\Delta=-{AB-CD\over A^2-C^2}, \qquad \Pi=-{AD-BC\over A^2-C^2}.\label{delta1}
\end{equation}
Multiplying (\ref{ventitre}) by $i\sigma_1^+\gamma^{\rho\theta}$ we obtain the consistency relation:
\begin{equation}
\Delta^2\,-\,\Pi^2=1. \label{eq1}
\end{equation}
The gaugino variation can also be cast in the form (\ref{ventitre}) with:
\begin{equation}
\Delta={4e^h+e^{-h}(a^2-1) \over 2a' }, \qquad \Pi=-{10e^h\lambda ' \over a' }, \label{delta2}
\end{equation}
so that from consistency of (\ref{delta1}) with (\ref{delta2}) we obtain other two equations.
Finally we have to take into account the $\psi_\rho$ variation, giving the $\rho$ dependence of the spinor and another equation:
\begin{equation}
[2+{1\over2}e^{-2h}(a^2-1)]\,\Delta-\,\partial_\rho\Pi=0. \label{eq4}
\end{equation}
One can verify that all these equations are solved by the functions in \cite{mn2}:
\begin{equation}
e^{2h}=\rho \coth{2\rho}-{\rho^2 \over \sinh^2{2\rho}}-{1\over4},\qquad
a={2\rho \over \sinh{2\rho}},\qquad
e^{10\lambda}={2e^h\over \sinh{2\rho}}.
\end{equation}
Note that this system consists of four equations for three functions and, although there is no evident relation between them, nevertheless a solution exists.
\section*{Appendix C: other solutions}
The previous equations are quite hard to solve due to their quadratic nature
induced by the non abelian gauge field.
In the abelian case $U(1)^+ \times U(1)^-$ with two or three scalars
the equations reduce to the linear ones in \cite{martelli,us}.
The next-to-simple example with quadratic equations other than the one by
Chamseddine-Volkov \cite{volkov} is with $SU(2)^+$, $\beta =0$,
and two scalars
($\lambda_1$ and $\lambda_2=\lambda_3$). The
four singlets of the inert $U(1)^-$ define a consistent truncation. They
transform in the $1+3$ representation of $SU(2)^+$. With an $SU(2)^+$
gauge transformation, we can rotate the triplet to its third component
$\lambda_2$. We can try to find a solution with only the fields in Appendix A.
Compared with the MN solution we have
an extra equation from the gaugino variation in the form (\ref{ventitre}) with:
\begin{equation}
\Delta=-{4e^h-e^{-h}(a^2-1) \over 4a}, \qquad \Pi={e^hd'\,\coth{d} \over 2a},
\end{equation}
for a total of six equations for four variables.
After the change of variables:
\begin{equation}
{du\over d\rho}=\cosh{d},\qquad d\equiv \lambda_1-\lambda_2, \label{cambiovariabili}
\end{equation}
 four equations
reduce to the MN ones. The two new gaugino equations give an equation
for $d$ and an equation which only depends on $a$ and $h$
($d$ cancels out)
and that, quite remarkably, is automatically satisfied by the MN solution.
Solving for $d$ we find:
\begin{equation}
d(u)=-{\rm arcsinh}\left({c_1\over 2\sqrt{1-8u^2-\cosh{4u}+4u\sinh{4u}}}\right), \label{su22sol}
\end{equation}
$c_1$ being an integration constant. This one doesn't flow to the MN
 solution, because in the IR $u\sim \rho^{1/3}$. We explicitly checked that the second order equations of motion are satisfied.
Using formulae in \cite{cve1}, we obtain the ten dimensional solution which presents a good type singularity (according to
the criteria of \cite{mn1}). The U.V. normalizable behavior of the solution indicates that it corresponds to the attempt of giving a VEV to scalar fields. Since these scalars are massive to begin with, we expect an instability in QFT that may explain the singular behavior of the supergravity solution.


\begin{thebibliography}{99}
\bibitem{vafa} M. Bershadsky, C. Vafa and V. Sadov,
Nucl. Phys. B463 (1996) 420; hep-th/9511222.
\bibitem{mn1} J. M. Maldacena and C. Nu$\tilde{\rm n}$ez,
Int. J. Mod. Phys. A16 (2001) 822;  hep-th/0007018.
\bibitem{lerche} A. Klemm, W. Lerche, P. Mayr, C.Vafa and N. Warner,
Nucl. Phys. B477 (1996) 746, hep-th/9604034.
\bibitem{mn2}J. M. Maldacena and C. Nu$\tilde{\rm n}$ez,
Phys. Rev. Lett. 86 (2001) 588; hep-th/0008001.
\bibitem{martelli} J. P. Gauntlett, N. Kim, D. Martelli
and D. Waldram, Phys. Rev. D64 (2001) 106008; hep-th/0106117.
\bibitem{us} F. Bigazzi, A. L. Cotrone and A. Zaffaroni,
Phys. Lett. B519 (2001) 269; hep-th/0106160.
\bibitem{jpp}C. V. Johnson, A. W. Peet and J.
Polchinski,
Phys. Rev. D61 (200) 086001; hep-th/9911161.
\bibitem{ks}I. R. Klebanov and M. J. Strassler, JHEP 0008 (2000) 052;
 hep-th/0007191;  C. P. Herzog, I. R. Klebanov and P. Ouyang, hep-th/0108101.
\bibitem{sonne} J. Sonnenschein and A. Loewy,
JHEP 0108 (2001) 007; hep-th/0103163.
\bibitem{sz} A. Salam and E. Sezgin,
Phys. Lett. B126 (1983) 304.
\bibitem{evans} J. Babington and N. Evans; hep-th/0111082.
\bibitem{seibergwitten} N. Seiberg and E. Witten,
Nucl. Phys. B426 (1994) 19; hep-th/9407087.
\bibitem{bgz} F. Bigazzi, L. Girardello and A. Zaffaroni,
Nucl. Phys. B598 (2001) 530; hep-th/0011041.
\bibitem{pp} A. W. Peet and J. Polchinski,
Phys. Rev. D59 (1999) 065011; hep-th/9809022.
\bibitem{peskin} M. E. Peskin; hep-th/9702094.
\bibitem{kw} I. R. Klebanov and E. Witten,
Nucl. Phys. B556 (1999) 89; hep-th/9905104.
\bibitem{gubser} S. S. Gubser, A. A. Tseytlin and M. S. Volkov,
JHEP 0109 (2001) 017; hep-th/0108205.
\bibitem{ds} M. R. Douglas and S. H. Shenker,
Nucl. Phys. B447 (1995); hep-th/9503163.
\bibitem{cve1} M. Cvetic, J. T. Liu, H. Lu and C.N. Pope,
Nucl. Phys. B560 (1999) 230; hep-th/9905096;
M. Cvetic, H. Lu and C.N. Pope,
Phys. Rev. D62 (2000) 064028; hep-th/0003286.
\bibitem{minasian}J. T. Liu and R. Minasian,
Phys. Lett. B457 (1999) 39; hep-th/9903269.
\bibitem{volkov} A. H. Chamseddine and M. S. Volkov,
Phys. Rev. Lett. 79 (1997) 3343; hep-th/9707176.
\end{thebibliography}
\end{document}